\documentclass{article}
\usepackage{ijcai24}

\usepackage{times}
\usepackage{soul}
\usepackage{url}
\usepackage[hidelinks]{hyperref}
\usepackage[utf8]{inputenc}
\usepackage[small]{caption}
\usepackage{graphicx}
\usepackage{amsmath}
\usepackage{amsthm}
\usepackage{booktabs}
\usepackage{algorithm}
\usepackage{algorithmic}
\usepackage[switch]{lineno}
\usepackage[acronym,shortcuts,nomain,style=super,nonumberlist,nopostdot]{glossaries}
\usepackage{graphicx}
\usepackage{subcaption}
\usepackage{cleveref}
\usepackage{array} 

\crefname{figure}{Figure}{Figures}
\crefname{table}{Table}{Tables}

\newacronym{ai}{AI}{Artificial Intelligence}
\newacronym{xai}{XAI}{Explainable Artificial Intelligence}
\newacronym{gqm}{GQM}{Goal Question Metric}
\newacronym{ml}{ML}{Machine Learning}

\newcommand{\eg}{e.g.}

\title{Perspectives on Explanation Formats From Two Stakeholder Groups in Germany: Software Providers and Dairy Farmers}

\iftrue
\author{
Mengisti Berihu Girmay $^1$
\and
Felix Möhrle $^1$\\
\affiliations
$^1$ RPTU Kaiserslautern-Landau, Kaiserslautern, Germany\\
\emails
mengisti.berihu@rptu.de, felix.moehrle@cs.rptu.de\\
}
\fi

\begin{document}

\maketitle

\begin{abstract}
This paper examines the views of software providers in the German dairy industry with regard to dairy farmers' needs for explanation of digital decision support systems.
The study is based on mastitis detection in dairy cows using a hypothetical herd management system.
We designed four exemplary explanation formats for mastitis assessments with different types of presentation (textual, rule-based, herd comparison, and time series).
In our previous study, 14 dairy farmers in Germany had rated these formats in terms of comprehensibility and the trust they would have in a system providing each format.
In this study, we repeat the survey with 13 software providers active in the German dairy industry.
We ask them how well they think the formats would be received by farmers.
We hypothesized that there may be discrepancies between the views of both groups that are worth investigating, partly to find reasons for the reluctance to adopt digital systems.
A comparison of the feedback from both groups supports the hypothesis and calls for further investigation.
The results show that software providers tend to make assumptions about farmers' preferences that are not necessarily accurate.
Our study, although not representative due to the small sample size, highlights the potential benefits of a thorough user requirements analysis (farmers' needs) to improve software adaptation and user acceptance.
\end{abstract}


\section{Introduction}

With the increasing advance of \ac{ai}, more and more applications are also emerging in agriculture.
In arable farming, digitalization has enabled the concept of precision farming, where application rates of inputs (\eg{} fertilizer) are precisely adjusted for different field zones.
Traditionally, digital systems use rule-based programming logic (\eg{} if-then-else).
However, with its growing popularity, it is to be expected that \ac{ai} will also be increasingly integrated into these practices.
For example, satellite or drone images can be analyzed using image recognition methods to monitor plant growth and optimize cultivation --- or to examine plants for diseases and pests in order to apply plant protection products at an early stage where necessary.
\ac{ai} also offers new possibilities for crop yield forecasts \cite{berger2020combining} or predictions of the expected water consumption \cite{linaza2021data}.
There is similar potential in livestock farming. Image-based or audio-based AI methods can be used to detect animal diseases (\eg{} lameness or mastitis) or adverse animal behavior (\eg{} tail biting).
Further applications are likely to emerge in the future, and it is to be anticipated that \ac{ai} will be a contributing factor to increasing the overall productivity of many farms \cite{hoxhallaripotential,agronomy11061227}.

While many believe that digitalization and, increasingly, \ac{ai} hold great potential for the future, the adoption of digital solutions in practice is rather cautious, even though many technologies have been on the market for many years \cite{dorr2022handbook}.
There are many studies that investigate possible influencing factors for the poor acceptance \cite{pfeiffer2021understanding,mohr2021acceptance,reichardt2009dissemination}.
Farmers continually express dissatisfaction about the high complexity of systems as well as poor user-friendliness.
Other factors include a lack of interoperability with other systems and concerns about data protection.

A recent study with crop farmers has shown that farmers tend to stick with their current level of technology, suggesting a low likelihood that they will adopt new information-intensive technologies \cite{gabriel2023adoption}.
Further research emphasizes that the explainability of digital systems plays a crucial role, as incomprehensible and non-transparent decision recommendations lead to a lack of trust among users \cite{10.1145/3529320.3529321,shin2021effects}.
With the advent of \ac{ai} in digital farming solutions, these adoption barriers are likely to rise even further.
This is not least due to the black-box nature of many \ac{ai} solutions, whose rationale is difficult for end users to understand (why is the system giving me this recommendation?).

A number of recent studies have addressed the need for better explainability of digital systems.
However, many researchers focus mainly on safety-critical use cases, \eg{} in the automotive industry or healthcare \cite{ahmad2018interpretable,glomsrud2019trustworthy}.
With regard to the agricultural sector, we believe that not enough research is being done on the needs of farmers and, as a result, many digital solutions are not optimally adapted to their needs in practice.

We see great potential for improvement particularly in decision support systems, which are designed to help farmers in their day-to-day decision-making.
An example is the assessment of animal health and the recommendation of treatments.
Without comprehensible reasoning, farmers may be reluctant to believe the system's judgement.
By providing a rationale (\eg{} rise in body temperature or prolonged sitting) in addition to recommendations, the assessment can be made more comprehensible to farmers to increase their willingness to trust a system.

Based on these observations, an entire branch of research has emerged over recent years, usually referred to as \ac{xai}.
The concept of \ac{xai} aims to make results generated by \ac{ai} more comprehensible and transparent.
In this respect, explainability can be considered a non-functional requirement that, if implemented appropriately, can contribute to better user experience.


Recent focus has mostly been on the development of new \ac{xai} techniques, with little attention being paid to evaluation with stakeholders.
Many authors, however, emphasize the need for stakeholder involvement to ensure meaningful benefits \cite{vermeire2021choose,make3030032}.
In the agricultural ecosystem, farmers and software providers are two key stakeholder groups, each with different interactions and expectations of \ac{ai} technologies \cite{9920137}.
Farmers are looking for directly actionable insights to make daily operational decisions.
On the other hand, software providers may focus more on the functionality and innovation of algorithms, often at a level of complexity that is not easily understood by non-experts.
This discrepancy in the perception of explainability can impair the benefits and acceptance for end users.


In our work, we explore the notion of explainability of decision support systems using a small case study (online survey) from the German dairy farming sector, which we conducted over a period of 6 weeks.
A hypothesis behind our study was that there could be discrepancies between the actual needs of farmers and the needs perceived by software providers, which could be one reason for the generally low acceptance of digital systems.
To this end, we asked software providers for their views on the needs of dairy farmers regarding the explainability of digital systems.
By software providers, we refer to employees of software development companies who are involved in the development of such systems.
The study was designed with a relatively small scope and therefore cannot provide representative results.
Rather, the intention was to provide initial findings and possibly a basis for future work.




\section{Related Work}

There are different views in literature on what the term explainability entails and how it should be defined.
We rely on the definition of \cite{atf2023human}, who refine explainability into two aspects, namely comprehensibility and trust.
Comprehensibility refers to how well the user is able to understand the reasons for a given fact (\eg{} a decision or classification made by \ac{ai}), while trust refers to the user's willingness to rely on that decision.

There is a lot of research in this field, although the focus is predominantly on two aspects.
One strand of research focuses on the explainability of models, which is particularly useful from the perspective of \ac{ml} engineers.
Another area is dedicated to user experience and examines the needs of end users in terms of explainability, \eg{} how they react to certain explanation methods.
There is, however, little research examining how explainability created during system development of real-world systems is received by end users to investigate possible discrepancies in the perspectives and needs of developers and end users.

The fact that software providers and end users do not necessarily have the same requirements in terms of explainability  was recognized, for example, by \citeauthor{preece2018stakeholders} \shortcite{preece2018stakeholders}.
The authors classify explainability based on four stakeholders, namely developers, theorists, ethicists, and users.
They have observed that users and ethicists focus more on validation and what a system does rather than how it is built, which is more of a concern for developers.

We believe that by better considering the needs of the end user when creating explainability during the development process, a contribution can be made to improving user experience and ultimately user acceptance.
This opinion is not new and is also expressed by other authors in literature.
\citeauthor{Miller2019} \shortcite{Miller2019} and \citeauthor{lipton2018mythos} \shortcite{lipton2018mythos} observe that users tend to place more trust in systems whose functions they can understand and emphasize the potential of explainability tailored to the end user.
\citeauthor{navarro2021desiderata} \shortcite{navarro2021desiderata}
and \citeauthor{9920137} \shortcite{9920137} criticize that the evaluation of explainability is largely theoretical and often lacks practical evaluation for real applications with end users.
\citeauthor{kohl2019explainability} \shortcite{kohl2019explainability} further state that explanations can help to improve user-friendliness and minimize human error, which will contribute to their user experience.

There are advances in many areas in which end users are increasingly involved in research into explainability.
\citeauthor{kuznietsov2024explainable} \shortcite{kuznietsov2024explainable} provide a systematic review and a framework for explainable methods for safe and trustworthy systems in the autonomous driving domain.
\citeauthor{SCHOONDERWOERD2021102684} \shortcite{SCHOONDERWOERD2021102684} present a human-centered design approach to develop explanations for \ac{ai} systems by involving experts and end users in the development process in the medical domain.
In the financial sector, \citeauthor{benhamou2021explainable} \shortcite{benhamou2021explainable} aim to improve the credit scoring system by making it better comprehensible and interpretable with \ac{xai}, whereby the evaluation is carried out with both functional and human involvement.

There are also a few examples in agriculture, such as a study by \citeauthor{ryo2022explainable} \shortcite{ryo2022explainable} that explores different interpretable machine learning tools for global maize yield data to improve the understanding of agricultural data.
\citeauthor{cartolano2024analyzing} \shortcite{cartolano2024analyzing} investigate the benefits of \ac{xai}, focusing on the comprehensibility of \ac{ml} models for end users using SHAP and LIME, but without direct involvement of farmers.
Overall, there is still a lack of research involving farmers and aligning explainability with their needs.
This is unfortunate as many decisions that farmers have to make on a daily basis affect the environment, animal welfare, and often involve considerable investments.
We therefore believe that improving decision support systems by making them easier for farmers to comprehend and increasing their trust can have great potential to improve farm operations.

\section{Research Goals and Design}

Our work builds on our previous study \cite{mengisti2024exploring}, which introduces a hypothetical herd management system that presents dairy farmers with assessments of mastitis in dairy cows.
Mastitis is a common infectious disease in dairy cattle and a major cause for the administration of antimicrobials.
Signs of an increased risk include a high number of somatic cells in the milk and an increased body temperature.
The target group of this study were farmers who were asked to evaluate four different explanation formats for susceptibility assessments. 
The participating farmers were asked how comprehensible they find each format and to what extent they are willing to trust a system that uses the respective format.

In this paper, we build on the aforementioned study and extend it by shifting the focus to software providers.
To this end, we reuse the hypothetical herd management system and the four explanation formats from the referenced work.
We define our research goal in accordance with \citeauthor{wohlin2012experimentation} \shortcite{wohlin2012experimentation} as follows.

\vspace{3mm}

\noindent
\begin{tabular}{|>{\rule{0pt}{\baselineskip}}p{.95\linewidth}|}
\hline
\textbf{Goal definition:} 
We analyze discrepancies in the perspectives of software providers and dairy farmers regarding the need for explainability.
We present software providers with four explanation formats for mastitis risk assessments that could be integrated into decision support systems.
We ask them about their expectations of how each format would contribute to farmers' comprehension and trust.
We compare their perceptions with the feedback we previously received from farmers in a separate study.\\
\hline
\end{tabular}

\vspace{3mm}

\begin{figure*}[tbh]
    \centering
    \includegraphics[scale=.9]{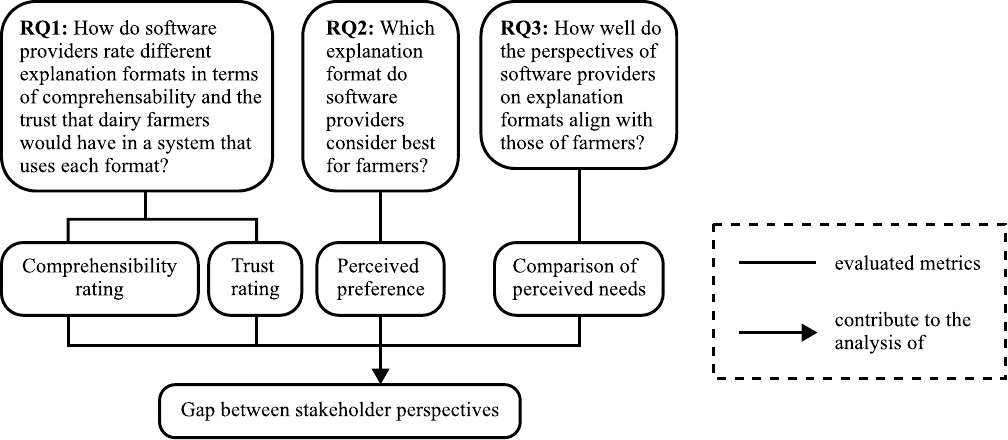}
    \caption{Research questions and related metrics}
    \label{fig:research-questions}
\end{figure*}

\noindent
Using the \ac{gqm} paradigm from \citeauthor{caldiera1994goal} \shortcite{caldiera1994goal}, we break down the objective of our study into three research questions and corresponding metrics, as shown in \cref{fig:research-questions}.
We present software providers with a hypothetical herd management system that assesses the risk of mastitis in dairy cows.
We designed the hypothetical system based on existing herd management systems in order to create a realistic use case.

Our first research question \textbf{RQ1} is how the software providers assess the suitability of the four exemplary explanation formats for farmers.
We assess suitability using two metrics: the comprehensibility of the formats for farmers and the trust that farmers would have in a system that provides this format.
Our second research question \textbf{RQ2} aims to determine a favorite among the four explanation formats from the software providers' perspective.
To this end, we ask them for their preference among the formats that they consider best suited for farmers.
Our third research question \textbf{RQ3} is whether there is a gap between the perception of software providers and that of farmers.
To this end, we compare the feedback from both stakeholder groups.

\subsection{Explanation Formats}

\begin{figure*}[h!tb]
    \centering
    \begin{subfigure}{0.5\textwidth}
        \centering
        \includegraphics[scale=.14]{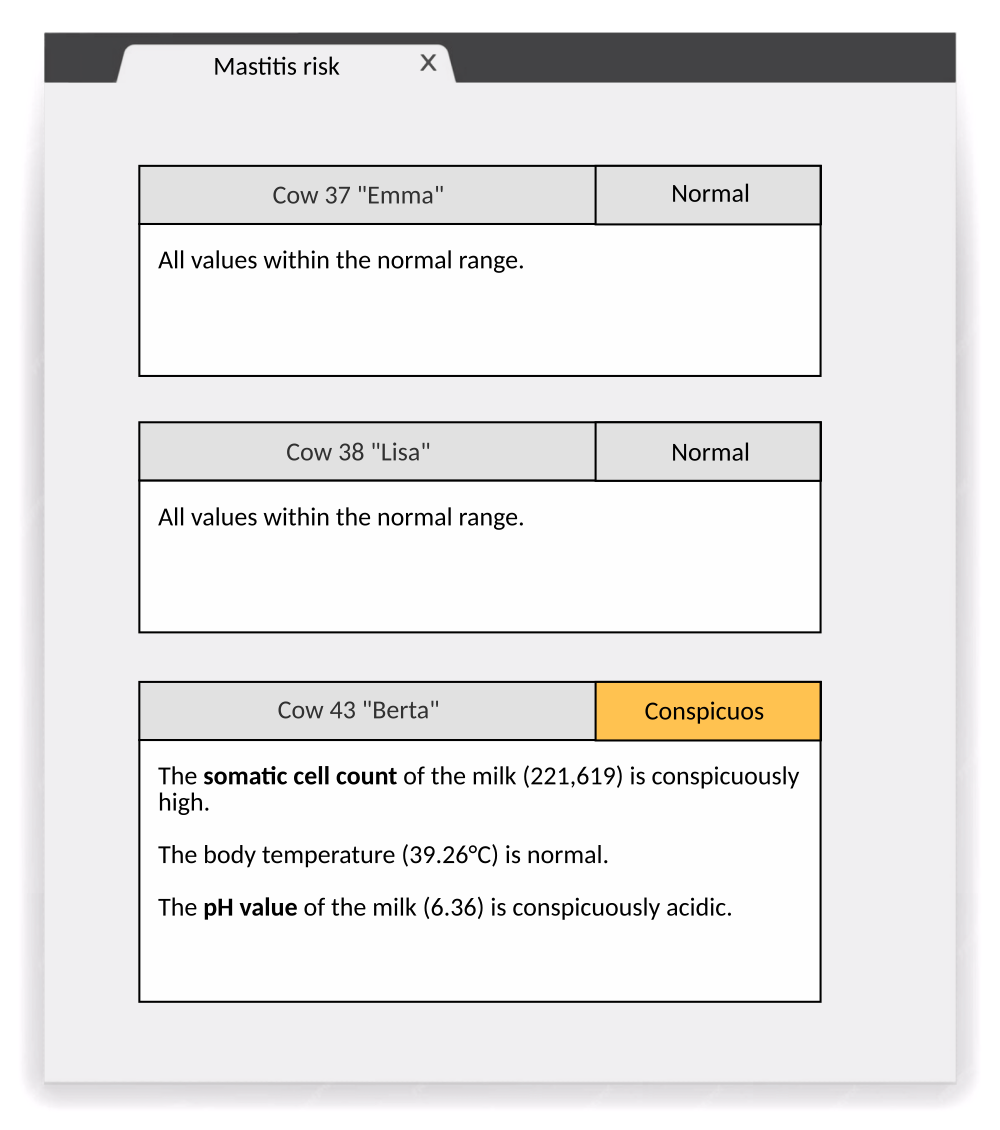}
        \caption{Textual explanation}
        \label{fig:textual}
    \end{subfigure}
    \begin{subfigure}{0.45\textwidth}
        \centering
        \includegraphics[scale=.14]{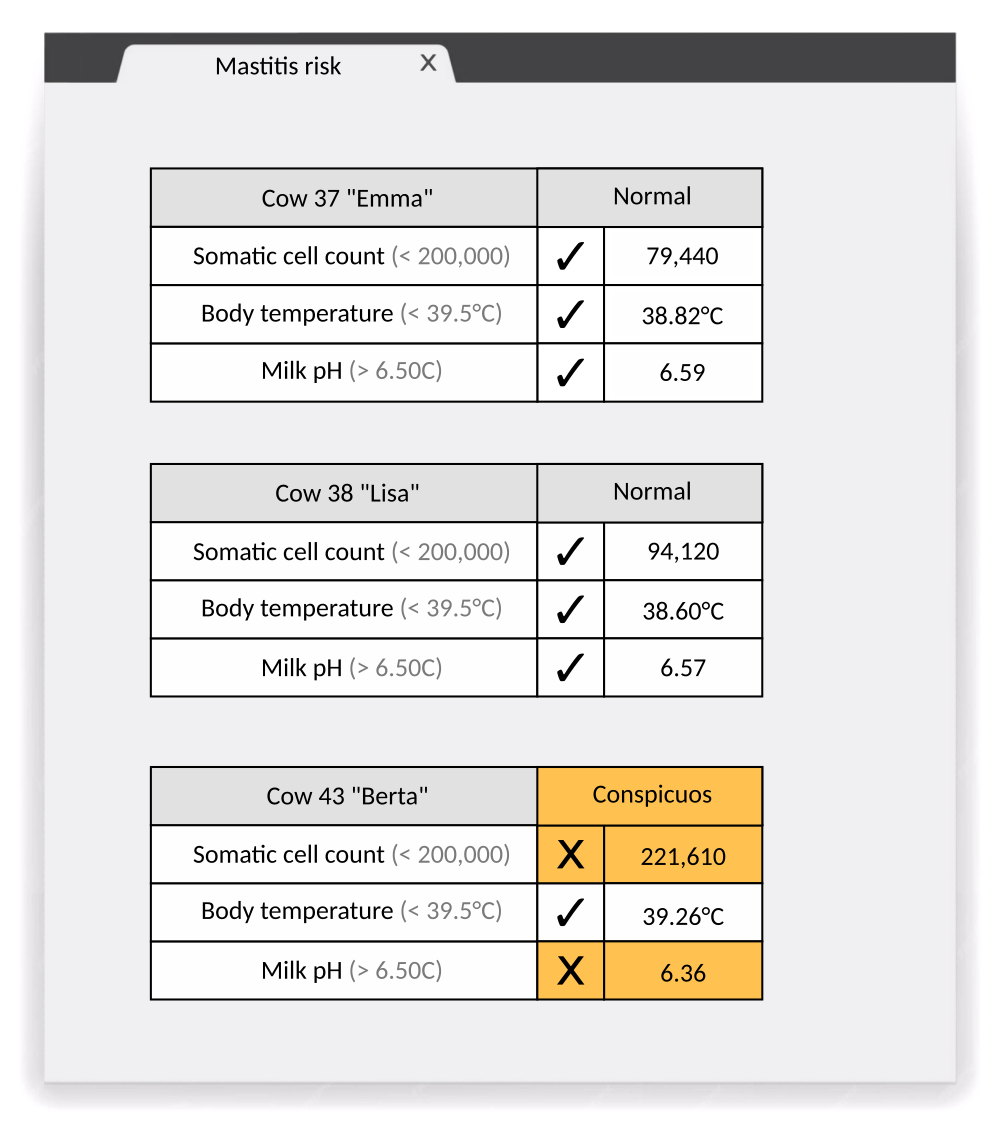}
        \caption{Rule-based explanation}
    \end{subfigure}
    
    \begin{subfigure}{0.5\textwidth}
        \centering
        \includegraphics[scale=.14]{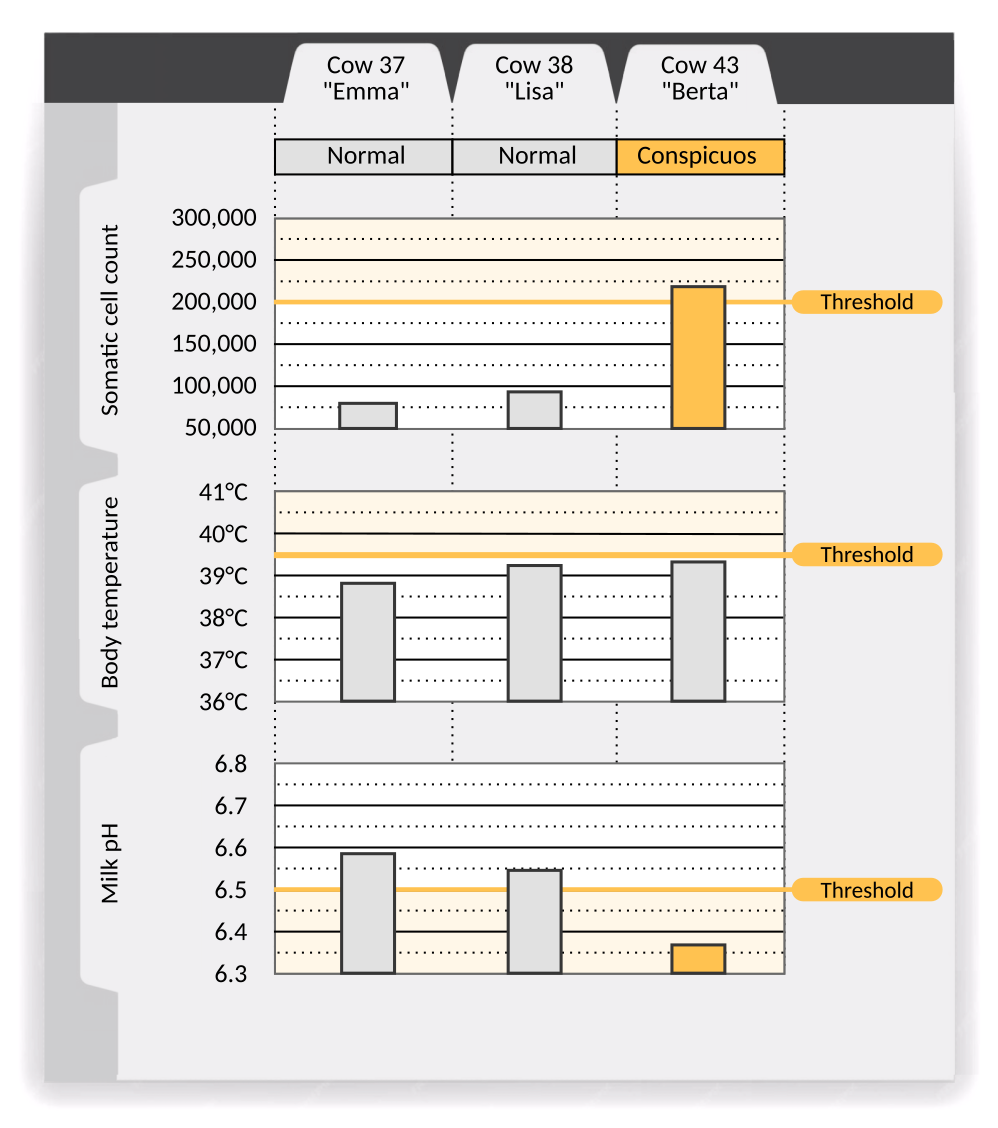}
        \caption{Herd comparison}
    \end{subfigure}   
    \begin{subfigure}{0.45\textwidth}
        \centering
        \includegraphics[scale=.14]{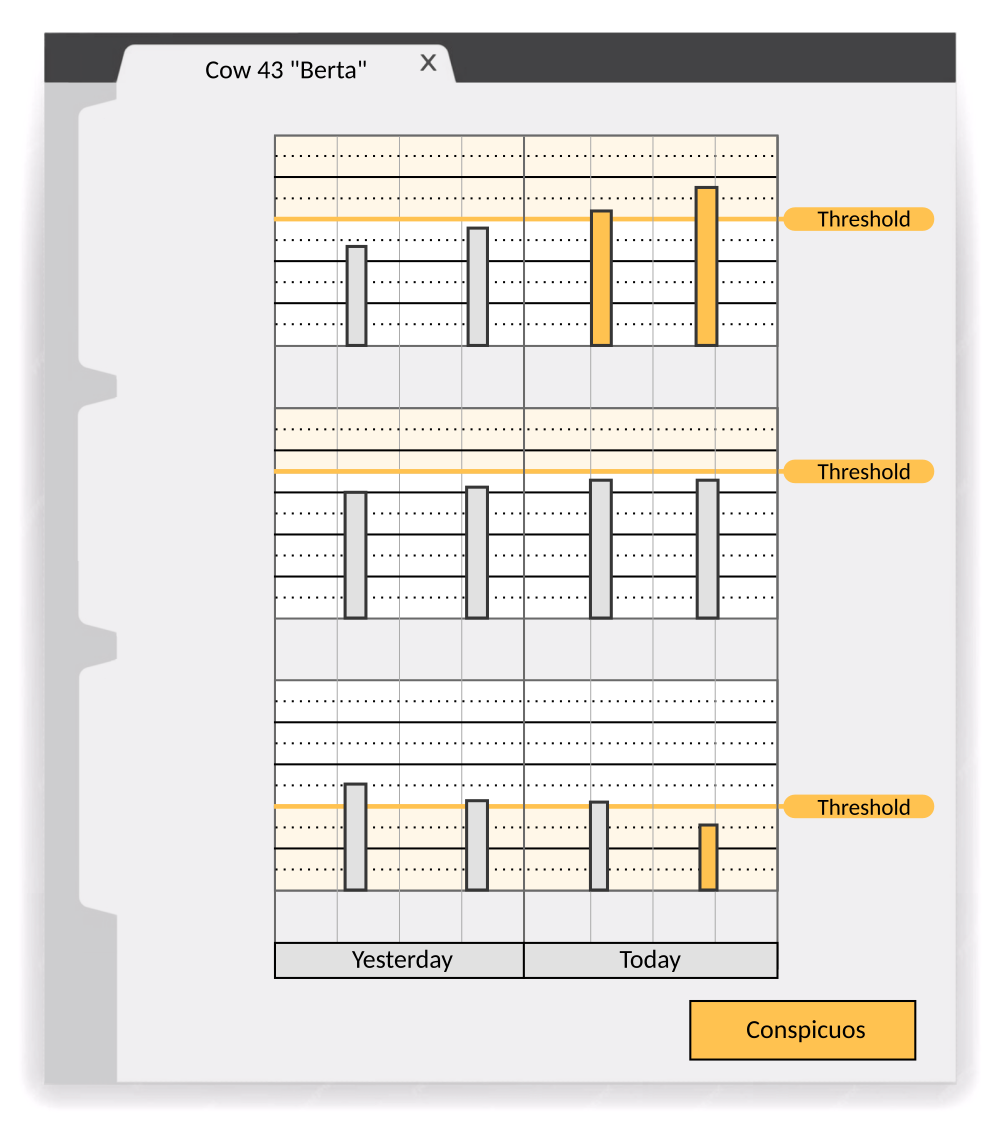}
        \caption{Time series}
    \end{subfigure}
    \caption{Four explanation formats (translated from the original German survey \protect\shortcite{mengisti2024exploring} into English)}
    \label{fig:explanation-formats-screenshots}
\end{figure*}

When designing the hypothetical herd management system and the explanation formats reused in our work, we originally followed the classification of \citeauthor{make3030032} \shortcite{make3030032}.
The authors classify techniques from the field of \ac{xai}, among others, by their scope (local, global) and their output format (numerical, rule-based, textual, visual).
From this classification, we extracted the classes textual, rule-based, global, and local, as these can be readily applied to the selected use case and transformed into realistic explanation formats.
In the textual format, we present assessments of the animals' health status as statements in natural language.
In the rule-based format, we provide rules that check whether the parameters of each individual cow are within a defined healthy range.
The global scope is realized by a herd comparison that shows and compares multiple cows.
The local scope is realized by a time series for each individual cow and shows the temporal changes of its parameters during the last measurements.
The four explanation formats are presented in \cref{fig:explanation-formats-screenshots}.

\subsection{Survey and Data Collection}

Recent studies on human-centered evaluation of user needs have shown that surveys are a good means of obtaining feedback on how well explanations perform in terms of comprehensibility and trust \cite{vanderWaa2021,doshi2017towards}.

Our work uses the tool LimeSurvey, in which we designed questions incorporating the metrics presented.
The survey contains four questions.
In the first question, the participating software providers rate on a 5-point Likert scale how comprehensible they think the four formats are for farmers (1~=~not at all comprehensible to 5~=~very comprehensible).
In the second question, they rate in the same way the trust they believe farmers would have in a system that uses each format (1~=~would not trust at all, 5~=~would trust completely).
In the third question, they choose which format they think is most suitable overall (single-choice).
In the fourth question, they justify their choice from question 3 with a free text entry.
Participation in the survey was anonymous.

We shared the survey with various associations and partners that offer \ac{ai}-based and other digital decision support systems in dairy farming.
Among them were three software companies that produce widely used herd management systems, which are particularly popular in Germany.
In doing so, we asked our contacts to forward the survey to suitable candidates in their networks.

\section{Results and Discussion}

In this section, we compare the perspectives of software providers and dairy farmers.
In this way, we wanted to determine to what extent their views align and whether there are discrepancies between both groups.

\subsection{Evaluation of Farmers' Feedback}

In our previous study, we asked 23 German dairy farmers for their opinions on the various explanation formats.
Of these, we evaluated the feedback from 14 who completed the survey in full.
Incomplete survey responses were discarded.

Many farmers found the time series preferable for decision-making and treatment selection.
Similarly, many farmers appreciated the herd comparison for its clarity and the fact that there are differences between herds, making the intra-herd comparison a particularly useful tool.
Overall, however, in the farmer survey, no format stood out significantly from the others.
All formats received positive and negative ratings, and for each format there were farmers who saw value in it.
The conclusion was that a good decision support system should offer several explanation formats and allow farmers to switch between them depending on the situation.
This preference was also expressed by several farmers in the free text responses.

\subsection{Evaluation of Software Providers' Feedback}

A total of 25 software providers took part in the survey conducted as part of this work.
Of their responses, 13 were complete.
We limited the analysis to the fully completed subset.

\subsubsection{Quantitative Analysis}

To answer our research question \textbf{RQ1}, we analyzed the ratings received by the software providers.
The quantitative analysis is only of limited use due to the small number of participants.
We carried it out nonetheless in the hope to identify possible tendencies.

\begin{figure}[tbh]
    \centering
    
    \begin{subfigure}{0.44\textwidth}
        \centering
        \includegraphics[width=\linewidth]{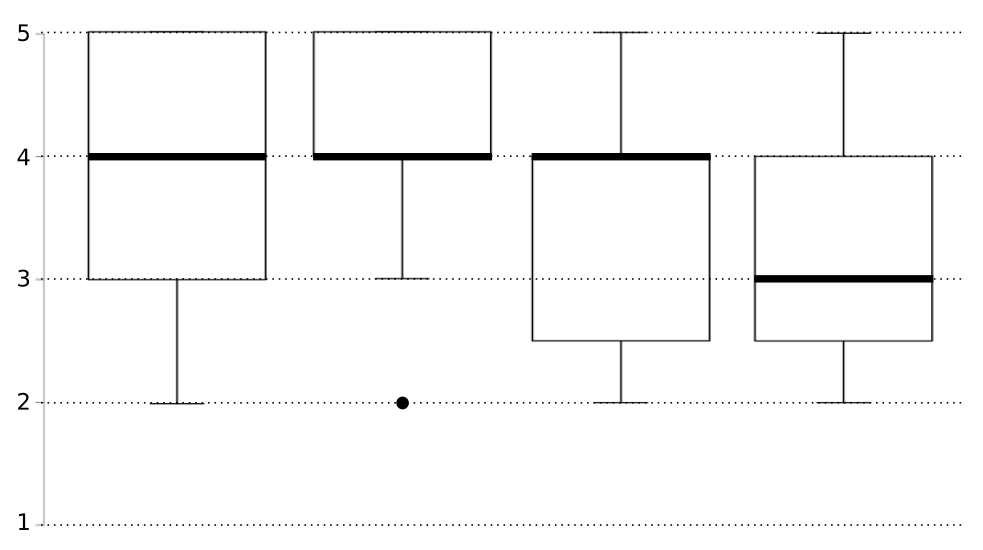}
        \caption{Comprehensibility ratings}
        \label{fig:textual}
    \end{subfigure}
    
    \begin{subfigure}{0.44\textwidth}
        \centering
        \includegraphics[width=\linewidth]{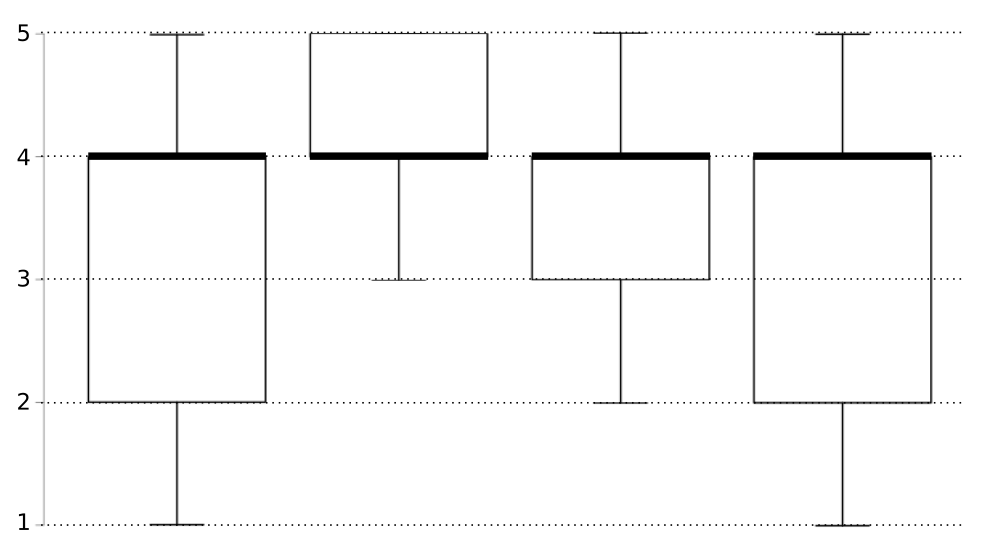}
        \caption{Trust ratings}
     \end{subfigure}
     
    \begin{subfigure}{0.35\textwidth}
        \centering
        
        \vspace{4mm}
        
        \includegraphics[width=\linewidth]{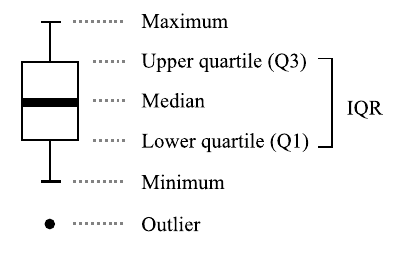}
        \caption{Legend}
     \end{subfigure}     
     
     \caption{Quantitative analysis results for the textual format, rule-based format, herd comparison, and time series (from left to right)}
    \label{fig:quantitative-results}
\end{figure}

The results are shown in \cref{fig:quantitative-results}.
Based on the participants' responses, we calculated the medians, lower and upper quartiles, minima and maxima of the comprehensibility and trust ratings (1 to 5) for each format.
The minima and maxima were calculated after removing outliers.
Outliers are values higher than the third quartile Q3 + 1.5 x IQR or lower than the first quartile Q1 - 1.5 x IQR with IQR being the interquartile range.

In the evaluation of the explanation formats' comprehensibility by software providers, the textual and rule-based format as well as the herd comparison received a median of 4.
The time series scored slightly lower with a median of 3.

In terms of trust, all explanation formats were rated similarly and received a median score of 4.
The rule-based format stood out slightly from the others with a higher minimum score of 3 (whereas the other formats had minimum scores of 1 and 2).

When voting for the best suited format for farmers according to \textbf{RQ2}, a tendency became apparent.
Almost half of the software providers opted for the rule-based format with 6 votes.
The textual format and the time series received 3 votes each.
The herd comparison only received 1 vote.

\subsubsection{Qualitative Analysis}

For the qualitative evaluation, we assessed the free-text responses of the software providers, who gave reasons as to why they preferred a particular explanation format.
It should be noted that not all participants gave a reason for their choice.

Of the software providers who preferred the textual format, two stated that this format is the most concise and displays the least amount of irrelevant information.
One of them added that in his/her experience herd managers have little time, making the textual format the best choice due to its time-saving nature with the least amount of unnecessary numbers and charts.

Regarding the rule-based explanation format, two software providers remarked that it is the easiest to implement and the clearest in terms of communication as rules are easy to understand for everyone.
Another participant noted that the format was preferable due to its clear structure and quick overview of all areas.

Of the participants who preferred the time series, one commented that it shows the development of critical values best, but that otherwise, if only recent warnings are to be reported, the rule-based format would be more suitable.
Another participant noted that for individual cows, some values may be slightly elevated on a permanent basis and therefore the development of the value is most important.
Another remark was that herd comparisons can quickly become confusing if too many animals are involved, which is why this participant prefers the time series --- but would find it best to be able to choose between several formats.

For the herd comparison format, the participants had conflicting opinions.
The participant who preferred this format stated that he/she considers it the clearest and most helpful for large herds.
Another participant, on the other hand, noted that it would be confusing with larger herds.

The feedback is summarized in \cref{tab:qualitative-summary}.

\begin{table}[tbh]
\centering
\begin{tabular}{lp{.61\linewidth}}
\toprule
Format      & Feedback \\
\midrule
Textual       & + Brief and concise information (2) \\
              & + Time-saving, least amount of unnecessary information (1) \\
              \\
Rule-based    & + Easy to implement, easy to understand and communicate (2) \\
              & + Clear structure and overview (1) \\
              & + Best for acute warnings (1) \\
              \\
Herd comparison & + Clearest format for large herds (1) \\
                & - Confusing for large herds (1) \\
                \\
Time-based    & + Shows development of critical values best (1) \\
              & + Copes best with unique characteristics of individual animals (1) \\
\bottomrule
\end{tabular}
\caption{Feedback of software providers on the explanation formats}
\label{tab:qualitative-summary}
\end{table}

\subsection{Comparison}

To answer our research question \textbf{RQ3}, we compared the feedback of both stakeholder groups.

The textual format was rated as well comprehensive by both groups (median 4.5 for farmers, 4.0 for software providers).
In terms of trust, it was also rated highly by both (median 4.0 both).
Among the farmers, it received a negative remark that it was not detailed enough and thus inferior to the more informative time series.
Some of the software providers, on the other hand, were of the opinion that farmers want information to be as concise and time-saving as possible, which indicates a disconnect in the perceived requirements.

The rule-based format was rated exceptionally highly by farmers (median 5.0) and high by software providers (median 4.0) in terms of comprehensibility.
It also scored well in terms of trust where it was rated equally by both groups (median 4.0).
The format was largely praised as being very clear by both farmers and software providers.
Among the software providers, there was not a single negative remark.
In the farmers' group, however, the format also received a negative comment by one participant who found it unclear, showing that, unlike the participating software providers, not all farmers are in favor of this type of presentation.

The herd comparison was rated the same by both groups in terms of comprehensibility (median 4.0).
In terms of trust, the farmers rated it slightly lower (median 3.5) than the software providers (median 4.0).
The majority of farmers responded positively, stating that it provides important information and that comparisons within a herd are most useful because each herd is unique.
In the group of software providers, some comments were contradictory.
One participant praised the format for its good clarity in large herds, while another criticized it for being unclear in precisely this case, indicating different perceptions of everyday practice on farms.

The time series received a high rating from the farmers in terms of comprehensibility (median 4.5), but a much lower rating from the software providers (median 3.0).
This indicates that farmers may be able to derive more value from the observed changes in the monitored health parameters (\eg{} body temperature) over time than the software providers.
In terms of trust, however, the farmers rated the time series very low (median 2.5), which raises a need for further clarification that unfortunately could not be resolved as part of the study.
In contrast, the software providers considered farmers' trust in the time series to be high (median 4.0), which reveals a different perception.
The majority of farmers praised the format because changes in an individual animal are most important for decision-making.
This assessment was also reflected in the comments from the software providers, indicating a similar perception.

With regard to the favored formats, the majority of software providers believed that the rule-based format is generally the best suited one for farmers.
In the group of farmers, however, the most favored format was the time series (the rule-based format followed in second place). 
A possible reason why the rule-based format was favored most among software providers could be that people involved with software artifacts (\eg{} code or specifications) tend to think in a rule-based way and may find this format particularly intuitive.
Among the farmers, however, not all participants had positive remarks towards the rule-based format, which indicates that the views between the two groups diverge slightly.

One finding from our small study is that the assumptions made by software providers about farmers (\eg{} desired scope of information, preferred type of presentation) are not always accurate.
Although our results are not representative enough to make universally valid statements, there is at least a discernible tendency that a better analysis of user requirements (farmers' needs) could enable better adaptation of the software and possibly increase user acceptance.

\section{Conclusion and Outlook}

In this paper, we examined the perspectives of software providers operating in the German dairy sector on the explanation needs of German dairy farmers.
As a use case, we used the detection of mastitis in dairy cows in a hypothetical herd management system.
In our previous study, 14 dairy farmers rated four explanation formats (textual, rule-based, herd comparison, and time series) for mastitis assessments in a survey according to their comprehensibility and the trust they would place in a system offering each format.
As part of this work, we repeated the survey with 13 software providers and asked them to rate how well the explanation formats would perform in the eyes of farmers.
We compared their opinions with the feedback previously gathered from the farmers and analyzed how well their opinions align.

The most favored format by farmers was the time series, while software providers considered the rule-based format to be best suited to farmers' needs.
The textual format was praised by the software providers for its conciseness (expecting that time saving is important for farmers), while some farmers remarked negatively that it contains too little information.
The rule-based format was mostly well received by farmers and was seen as the best choice by software providers, although some farmers found it unclear.
The herd comparison was generally well received by farmers, but received contradictory ratings from software providers.
The time series was the most favored format for being detailed and actionable by farmers, while being slightly less favored by software providers due to its complexity, suggesting that farmers may derive greater benefit from it.

A concluding observation from our small study is that the software providers' assumptions about farmers' preferences, such as the desired amount of information and preferred presentation style, are not always reflected in reality. 
Even though the representativeness of our results is limited, it seems reasonable to assume from our findings that a more thorough analysis of user requirements (farmers' needs) could help to better adapt software to farmers' needs and possibly increase user acceptance.

In future work, user persona studies could help to understand farmers and their requirements at a more refined level.
Such research could take into account demographic characteristics, previous experience with technology, and other factors.
In this way, software systems could be provided with user profiles tailored to different personas in order to present results in a convenient form and improve acceptance.

\appendix

\section*{Acknowledgments}

This research was supported by the Carl-Zeiss Foundation under the Sustainable Embedded AI project (P2021-02-009) and the German Federal Ministry for Economic Affairs and Climate Action under the NaLamKI project (01MK21003[A-J]).

\bibliographystyle{named}
\bibliography{literatures}

\end{document}